\documentclass [12pt,a4paper,draft]{article}

\usepackage{times}

\DeclareFontFamily{OT1}{times}{}
\DeclareFontShape {OT1}{times}{m }{n }{ <-> ptmr }{}
\DeclareFontShape {OT1}{times}{bx}{n }{ <-> ptmb }{}
\DeclareFontShape {OT1}{times}{m }{it}{ <-> ptmri}{}
\DeclareFontShape {OT1}{times}{bx}{it}{ <-> ptmbi}{}
\begin{document}

\title{\vspace{-1.cm} \bf Depleted-Uranium Weapons:\\ the Whys and Wherefores}

\author{André Gsponer\\
\emph{Independent Scientific Research Institute}\\ 
\emph{ Box 30, CH-1211 Geneva-12, Switzerland}\\
e-mail: isri@vtx.ch\\
~\\
{\bf Postface to a book}\\ {\bf to be published by the Bertrand Russell Foundation}}

\date{ISRI-03-03.25 ~~ \today}

\maketitle

\begin{abstract}

It is recalled that the only non-nuclear military application in which depleted-uranium alloys out-perform present-day tungsten alloys is long-rod penetration into a main battle-tank's armor.  However, this advantage is only on the order of 10\%, and it disappears when the uranium versus tungsten comparison is made in terms of actual lethality of complete anti-tank systems instead of laboratory-type homogeneous armor steel penetration capability.

  Therefore, new micro- and nano-engineered tungsten alloys may soon be at least as good if not better than uranium alloys, enabling the production of tungsten munition which will be better than existing uranium munition, and whose overall life-cycle cost will be lower, due to the absence of the problems related to the radioactivity of uranium.

The reasons why depleted-uranium weapons have been introduced into arsenals and used in Iraq and Yugoslavia are analysed from the perspective that their radioactivity must have played an important role in the decision making process. It is found that depleted-uranium weapons belong to the diffuse category of low-radiological-impact nuclear weapons to which emerging types of low-yield (i.e., fourth generation) nuclear explosives also belong.

It is concluded that the battlefield use of depleted-uranium in the 1991 Gulf War, which broke a 46-year-long taboo against the intentional use or induction of radioactivity in combat, has created a military and legal precedent which has trivialized the combat use of radioactive materials, and therefore made the use of nuclear weapons more probable.

\end{abstract}

\newpage 	

{\bf \Large Contents}

\begin{tabular}{llr}

    &                                                                &    \\
1   & {\bf Introduction}                                             &  3 \\
    &                                                                &    \\
2   & {\bf Are uranium alloys really better than tungsten alloys for}&    \\
    & {\bf conventional weapons?}                                    &  5 \\
    &                                                                &    \\
2.1 & Penetration aids for bombs or cruise missiles                  &  7 \\
2.2 & Shield plates for armored tanks                                &  8 \\
2.3 & Penetrators for anti-tank weapons                              &  9 \\
2.4 & Munitions for ground-attack aircrafts or close-in-defense guns & 14 \\
    &                                                                &    \\
3   & {\bf Are depleted-uranium weapons conventional or nuclear}     &    \\
    & {\bf weapons?}                                                 & 15 \\
    &                                                                &    \\
3.1 & Nuclear and conventional materials                             & 16 \\
3.2 & Nuclear and conventional weapons                               & 18 \\
3.3 & Depleted-uranium and fourth generation nuclear weapons         & 20 \\
    &                                                                &    \\
4   & {\bf Discussion and conclusion}                                & 22 \\
    &                                                                &    \\
    & {\bf Acknowledgments}                                          & 27 \\
    &                                                                &    \\
    & {\bf References}                                               & 27 \\

\end{tabular}

\newpage

\section{Introduction}

When conventional weapons containing depleted-uranium were used for the first time in combat during the 1991 Gulf War, there was immediate reaction from the general public, environmental groups, and anti-nuclear activists.  This reaction has now turned into a major environmental, medical, technical, legal, and political opposition with many thousands ``anti-depleted-uranium activists'' in many countries striving to outlaw the use of depleted-uranium in weapons and consumer goods.

What drove this reaction was common sense.  Indeed, while there had been little  public or parliamentary opposition to the development and deployment of conventional weapons containing depleted-uranium before the Gulf War, the immediate consequence of their \emph{actual use} in Iraq was the realization that uranium (whether natural, enriched, or depleted) is not a trivial material that can be placed in the same category as tungsten or steel.  In fact, it was this same commonsense understanding of the deep political significance of the military use of a radioactive material that drove the only parliamentary reaction, by Senator Bob Dole, to the 1978 decision by the U.S.\ Department of Defense to use depleted-uranium for making bullets \cite{DOLE-1978-}.

Of course, an important reason for the lack of significant early opposition to the military use of depleted-uranium is that its radiological impact is very much lower than that of existing types of nuclear weapons:  atomic and hydrogen bombs.  As is well known, depleted-uranium is only about half as radioactive as natural uranium, which is a low radioactive material.  But being radioactive means that any uranium-based material is \emph{qualitatively} different from any non-radioactive material, and therefore means that any use of uranium has important medical, technical, legal, and political implications.  Moreover, there is no doubt that these differences have always been perfectly understood in professional circles, as is witnessed by the considerable amount of legislation dealing with all types of radioactive materials, and the special efforts that had to be made, for example in the United States, to release depleted-uranium for commercial use \cite{AEC--1971}, and to enable it to be incorporated in ``conventional'' weapons and possibly exported to foreign countries \cite{ISDCA1980-}.

The only context in which the use of depleted-uranium could have been ``benign'' is that of a nuclear war.  This is why many people perceived armor-piercing munitions containing depleted-uranium as a tolerable part of a strategy to deter a massive tank attack by the nuclear-armed Warsaw Pact Organization.  But Iraq in 1991 did not have nuclear weapons.  This made the first use of depleted-uranium weapons during the 1991 Gulf War, which broke a 46-year-long taboo against the intentional use or induction of radioactivity in combat, particularly shocking.

It is therefore understandable that there have been a lot of over-reactions.  On the one side many opponents predicted apocalyptic consequences on the environment and the affected populations, and on the other side many governmental and official bodies counter-reacted by excessively down-playing the consequences of the use of depleted-uranium weapons.

There have been many papers, both in newspapers and in professional journals, discussing the near- and long-term environmental, medical, and radiological consequencies of depleted-uranium weapons.  On the other hand, there have been surprisingly few published investigations on the whys and wherefores of depleted-uranium weapons --- namely the technical and military reasons why they were first introduced into arsenals, and the strategic-political reasons why they were then used against Iraq and former Yugoslavia, two non-nuclear-weapon states. 

In particular, it appears that there is no published critical study of the much-vaunted superiority of uranium-based over tungsten-based anti-tank wea\-pons, despite the fact that several major countries, most prominently Germany, have equipped their tank fleets with tungsten-based anti-tank weapons which are claimed to be good enough to defeat the armor of all existing tanks.  In fact, as will be seen in section 2, this paradox is easily dismissed by looking at the professional literature, which shows that while penetrators made of tungsten alloys are somewhat inferior to depleted-uranium ones, the overall performance of tungsten-based anti-tank weapons is not worse than that of their uranium-based counterparts.

Moreover, it appears that there is no published critical study that focuses on the only fundamental property that distinguishes uranium from its competitors, namely the fact that it is radioactive, and examines it from the point of view of its strategic and political consequences.\footnote{The same remark applies to the pyrophoric property of uranium: there appears to be no published study discussing the quantitative importance of this effect in tank warfare \cite{FIALK2003A}.} In effect, the numerous studies in which radioactivity is taken into account are only those dealing with the environmental, medical, and legal consequences of the combat-use of depleted-uranium.  The only exception seems to be a study in which the radiological effect due to the large-scale battlefield use of depleted-uranium is compared to that of existing and hypothetical fourth generation nuclear weapons \cite{GSPON2002-}.  This study will be reviewed in section 3, and put into perspective with other elements which suggest the conclusion that possibly the most powerful institutional force behind the development and deployment of depleted-uranium weapons must have been the ``nuclear lobby,'' which in all nuclear-weapon states is pushing towards the trivialization of the use of nuclear materials of all kinds, and of nuclear weapons of current or future types.

In this paper, therefore, an attempt will be made to understand the ``whys and wherefores'' of depleted-uranium weapons --- the technical and military advantages which could have justified their introduction into arsenals, and the political and strategic reasons which could explain their actual battlefield use, despite the environmental, medical, legal, and political drawbacks which were known long before they were developed and used.  In section 4, this attempt will be merged with what is already better-known about depleted-uranium weapons, and a conclusion derived to give additional arguments for an immediate ban of depleted-uranium weapons of all kind.

\section{Are uranium alloys really better than
         tungsten\\ alloys for conventional weapons?}

In a number of civilian and military applications the decisive physical characteristic of a material is its density.  The next decisive parameter is usually cost, which means that the material should be made of an element that is reasonably abundant on Earth.  If the density has to be as high as possible, the choice must be between four elements, namely tungsten, uranium, tantalum, and hafnium, which have as maximum density 19.2, 19.0, 16.7, and 13.3.  By comparison,  in crystal rocks, these elements are all about ten times less abundant than lead, density 11.3, which itself is about thousand times less abundant than iron, density 7.8.

The main technical reason why high-density materials are important in conventional weapons comes from the so-called ``Root Density Law,'' a very simple result that can be derived as an exercise by students of a final-year high-school physics class.  It says that when a rod of length $L$, made of some material of density $\rho_1$, penetrates at very high velocity another material of density $\rho_2$, the maximum penetration length under ideal conditions is $\sqrt{\rho_1/\rho_2} \times L$.  By ``ideal conditions'' it is meant that the two materials interact so fast and violently at their contact interface that they immediately melt, which is why these conditions are named ``hydrodynamic limit.'' However, even below this limit, the Root Density Law is pertinent for comparing processes where kinetic energy is exchanged between materials of different densities.

For example, in the case of a tungsten or uranium anti-tank penetrator of length $L$, the absolute maximum penetration depth into homogeneous steel is $\sqrt{19/7.8} \times L \approx 1.5 \times L$.  This means that the penetration power into homogeneous steel armor is at most 50\% higher for a tungsten or uranium rod than for a steel rod.  Therefore, in all applications where the Root Density Law applies, and where steel is replaced by a higher-density material, the potential benefit is at most 50\%.  Moreover, if instead of comparing the differences between the use of a high-density versus a low-density material, one compares the differences between the use of one high-density material with that of another high-density material (i.e., an uranium alloy versus a tungsten alloy), the differences to be expected from the unavoidable disparities in the respective mechanical properties of the two alloys will necessarily be smaller than 50\% in any application where the  conditions of use are close to the ``hydrodynamic limit.'' In other words, when one is discussing the merits and demerits of using an uranium instead of a tungsten alloy, one is in fact discussing small differencies, on the order of 10--20\%, not large differences which could turn an uranium-based weapon into a ``superweapon'' compared to the same weapon made out of tungsten!

Therefore, the whole discussion in this section turns around the impact of relatively small mechanical and metallurgical differencies between various alloys based on either uranium or tungsten.  Indeed, if these two elements had the same physical and chemical properties, and would therefore form similar alloys when combined with other elements, the obvious choice would be tungsten --- and in most cases uranium would immediately be discarded because of its radioactivity.  The problem is that in the conditions where the disparities in mechanical properties of different alloys have a visible impact, i.e.\ below the hydrodynamic limit, there are many concurrent processes in which each of these disparities has a different effect:  if one property is favorable in some process, it will not necessarily be so in another one.  Therefore, there is no alloy, based on either uranium or tungsten, which has at the same time all the desirable properties.  The result is that the choice of a given alloy is a compromise, and this is reflected in the fifty years long history during which either tungsten or uranium alloys have been considered as the ``best'' for one weapon's use or another. (For the history of their use in anti-tank weapons, see \cite{LANZ-2001-, ROSTK2000-, DAVIT1980-, SPEER1943-}.  For a number of general and technical references on tank warfare see \cite{GSPON1984-, SAHIN1985-, HARRI1986-}.)

To understand where we stand today, we have to review the major conventional weapons applications in which high-density materials are used:

\begin{itemize}

\item Penetration aids for bombs or cruise missiles;

\item Shield plates for armored tanks;

\item Munitions for anti-tank weapons;

\item Munitions for ground-attack aircrafts or close-in-defense guns.

\end{itemize}

These four sets of applications are discussed in the next sub-sections, which are organized in a logical sequence --- the information given in each one is useful to understand the next sub-sections.

\subsection{Penetration aids for bombs or cruise missiles}

In these applications, the high-density material is used to give more weight and a slimmer profile to the weapon, for example a free-fall or rocket-assisted bomb, so that it will have more kinetic energy per unit frontal area when hitting the target (an airport runway, a bunker, etc.).  This enables the warhead to penetrate deeper into the target before the explosive is detonated by a delayed fuse, which enormously increases the damage.  Obviously, in such applications it is mandatory that the nose of the penetrator, as well as the casing enclosing the warhead, remain functional as long as possible.  Their strength and hardness therefore need to be as high as possible.  From all that is known about the respective mechanical properties of uranium and tungsten alloys, the first choice is then clearly tungsten.\footnote{Moreover, the potential advantage of uranium over tungsten alloys exists only for penetrating monoblock metallic-targets, not for penetrating concrete or rocks.} 

In the case of long-range cruise missiles, the same reasoning applies, with the difference that weight considerations have a different impact on the design.  This is because a cruise missile is both a delivery system and a bomb.  If too much weight is given to the warhead, the range of the cruise missile will be considerably shortened.  Moreover, there is a an obvious trade-off between the amount of high-explosives that is carried and the weight of the penetration aid.  Either the conventional warhead is made as powerful as possible (or eventually replaced by a nuclear explosive) and the weapon is detonated on the surface or at a shallow depth into the target, or else some quantity of high-explosive is sacrificed to increase the weight of the penetrator in order to gain a factor which may be as large as ten in the case of a deep penetration.  The trouble is that the impact velocity of a cruise-missile is not as large as that of a bomb.  The impact conditions are even likely to be well below the ``hydrodynamic limit,'' so that little benefit is to be expected from the Root Density Law.  A hardened steel or titanium penetrator could therefore be just as good as a heavy-metal penetrator, and the possibility that the cruise missiles used in the Gulf War or in Yugoslavia carried a heavy-metal penetrator is only a speculation since ``there is no publicly available information at all that supports such an assumption'' \cite[p.165]{LILIO1999-}.

In summary, the best material currently available for the nose or casing of a hard-target penetrating warhead is tungsten heavy-alloy. However, if the material is used only as a ballast to increase the total weight of the warhead for better penetration performance, any heavy-material can be used in principle.

\subsection{Shield plates for armored tanks}

In the case of this application, it is important to know that the shield of a modern battle-tank is designed to protect the crew and the main components (engine, gun, ammunition, fuel-tank) from a large range of threats: conventional and nuclear bombs, shaped-charges delivered by bazookas or precision-guided munitions, anti-tank rounds fired by enemy tanks or heavy guns, landmines, etc.  For these reasons the shield of such a tank is a complicated structure comprising several layers of materials of different densities, interspaced with composite or fibrous materials, and possibly covered with reactive components such as high-explosive charges which automatically detonate to interfere with or destroy the darts or penetrators of anti-tank weapons. 

If one focuses on anti-tank rounds fired by enemy tanks or heavy guns, also called ``kinetic energy projectiles,'' the decisive element of a tank shield is that the main structural material, a thick layer of armor steel, is supplemented by one or two other layers, or arrays of plates, made of some tough material able to deflect, bend, break, or at least slow down, the anti-tank projectiles so as to prevent them from fully penetrating the main armor.  The effectiveness of this shielding mechanism has been demonstrated in many tanks built in the past decades.  From recent publications it appears, for example, that a long-rod penetrator emerging from a 1 cm thick steel plate hit at high obliquity has a residual penetration capability of only about 50\% of that of the original rod \cite{YAZIV2001-}.  This shows that the ``obliquity advantage'' of designing armored vehicles with glacis and sloped walls is quite important, and explains why anti-tank weapons capable of penetrating more than 60 cm of homogeneous armor steel are needed in order to defeat the armor of modern tanks \cite{LANZ-2001-}.

Therefore, an array of relative thin steel plates, put at an oblique position in front of the main armor, has a considerable shielding effect.  This effect can be further enhanced by the use of ``sandwiches'' in which steel plates are combined with layers of plastic or ceramic materials, or by replacing the steel with some higher-density material.  This calls for a hard high-strength heavy alloy, with tungsten the first choice again, even though the U.S.\ incorporated steel-clad depleted-uranium plates in the M-1 tanks deployed in Iraq during the Gulf War.  However, contrary to the first two applications in which the ``Root Density Law'' could bring an improvement of up to 50\%, the incorporation of heavy alloys in armor has only a marginal effect on the overall performance of the shield, especially if the heavy-material plates are made thinner than the original steel plates to keep the total weight constant.  Moreover, the difference between using uranium instead of tungsten will have an even more marginal effect, because there is no mechanical property of importance in this application that is very significantly different between alloys of either of these two materials.

\subsection{Penetrators for anti-tank weapons}

In this application, the discussion is more complicated because a number of distinct physical phenomena, which imply conflicting requirements on correlated mechanical properties, have to be taken into account.  For instance, anti-tank penetrators are first accelerated in a gun, where they are submitted to very high stress, vibrations, etc., until they exit from the gun muzzle where they violently separate from the ``sabot'' that was holding them during acceleration; secondly, they travel through air, where they can break because of bending or buckling;  and thirdly they eventually encounter a tank shield, where everything is done to defeat their action.  These three are, in gunnery jargon, the internal, external, and terminal ballistics problems --- to which one has to add the lethality problem, because piercing an armor is not enough to inflict damage or loss of function behind the armor.  Therefore, whereas in the applications considered in the two previous sub-sections the problem naturally led to a single set of requirements to be met by the properties of the material, and thus to relatively unsophisticated uranium or tungsten alloys, in the tank armor penetration application no single alloy has simultaneously the ideal properties suiting the requirements of the three ballistic problems.  For this reason, over the past fifty years, small differences between successive generations of uranium and tungsten alloys have resulted in either the former or the latter being perceived as ``better or worse'' for one ballistic phase or another, or for the anti-tank ballistic problem as a whole. 

To discuss these problems it is best to start from the target and work backwards to the gun, and then add overall performance and lethality considerations at the end.  Moreover, in the perspective of assessing the possible advantages of the use of uranium instead of tungsten, we will concentrate on the long-rod penetrators that are fired by the 120mm guns which today equip most of the main battle-tanks in service in the major Western powers's armies. Typically, these penetrators are heavy-material rods of about 60 cm in length and 2 cm in diameter.  Therefore, according to the Root Density Law, the maximum penetration depth should be about $60 \times 1.5 = 90$ cm in homogeneous armor steel, while the observed maximum is only about 60 cm for penetrators shot by a tank gun.  This is because the impact velocities of tank-fired projectiles are in the range of 1200 to 1700 m/s, which corresponds to a transition region somewhat below the hydrodynamic limit that is reached only in laboratory experiments where velocities of about 4000 m/s are achieved \cite[Figure 2]{SUBRA2001-}.

The slow evolution in heavy-metal alloy physics and metallurgy which led to such a penetration capability is described in a number of papers, for example references \cite{ROSTK2000-, DAVIT1980-} which emphasize depleted-uranium alloys and the 1950-1980 period, and reference \cite{LANZ-2001-} which emphasizes tungsten alloys and the more recent period.  In this evolution, the yardstick for comparing various alloys has become the depth of penetration in ``Rolled Homogeneous Armor'' (RHA) steel, i.e., the best possible armor steel.  This gives a very telling and impressive figure of merit, which is however only \emph{one} parameter of importance in the anti-tank ballistic problem.  Nevertheless, without entering into the historical details, whereas tungsten alloys were in general favored from the late 1950s until the early 1970s, the preference started to shift in the United States towards uranium alloys in the mid-1970s after the successful development of a new uranium alloy containing 0.75\% per weight titanium \cite{ROSTK2000-}.  This alloy overcame several technical and manufacturing problems that previously gave the preference to tungsten.  Moreover, the penetration capability of this U-3/4Ti alloy in RHA steel was consistently better than any tungsten alloy of equal density.  This improvement was the more impressive that the U-3/4Ti holes into RHA steel were both deeper and narrower than those produced by tungsten alloys, all other things been equal.

At the beginning, of course, the exact numbers showing how much the new uranium alloy was better than the tungsten alloys available at the time were classified (see, e.g., \cite{DAVIT1980-}).  But since at least 1990 numerous unclassified professional publications are given them, e.g., \cite{MAGNE1990-, GIAT-1994-, FARRA2001-}.  There are also numerous publications in which the overall average advantage of uranium alloys over tungsten alloys is given as a single number.  For example, S.P. Andrews et al., working on contract for the U.S.\ Army, claim that ``the ballistic performance of DU alloys and tungsten alloys against monolithic semi-infinite steel targets is similar --- at best 5 to 10\% difference'' \cite[p.255]{ANDRE1992-}.  Similarly, R.J. Dowding of the U.S.\ Army Research Laboratory at the Aberdeen Proving Ground, states that ``depleted-uranium alloys outperform tungsten alloys by 10\% at ordnance velocities'' \cite[p.4]{DOWDI2000-}.

Looking at the data, one finds as expected that the penetration advantage of uranium over tungsten alloys is less pronounced for high velocity projectiles because the conditions are closer to the hydrodynamic limit.  For instance, comparing the penetration efficiencies at 1200 and 1700 m/s impact velocities, the advantages are 25 and respectively 5\% in reference \cite[Fig.3]{MAGNE1990-}, 20 and 15\% in reference \cite[p.43]{GIAT-1994-}, and finally 25 and 15\% in reference \cite[Fig.2]{FARRA2001-}.  But these numbers should be taken with caution.  While they clearly demonstrate an advantage of about 10 to 20\% over the velocity range of interest for anti-tank weapons, they derive from drawings that are quite poor according to normal scientific standards: there are no error bars and the figures are clearly drawn to impress the non-specialists (see, especially, reference \cite{GIAT-1994-}).  Therefore, since these data constitute the main (if not \emph{only}) objective element that supports the thesis that uranium alloys are more effective than tungsten alloys for anti-tank weapons, it is important to stress that this relatively small effect is at the origin of much of the depleted-uranium controversy, because most of the depleted-uranium expended in Iraq was precisely in the form of anti-tank penetrators \cite{LANZ2003-}. 

A second interesting feature shown by the data, e.g., reference \cite{MAGNE1990-}, is that the penetration dynamics of uranium rods is different from that of tungsten rods with the same dimensions and weight.  Normally, when a rod of some material penetrates at high velocity into a target material, a mushroomed region forms near the impacting end, and its nose erodes (as the penetrator burrows into the target) by giving up material that is ``back-extruded'' from the penetrator-target interface while the interface moves forward into the target.  This process is explained and illustrated in many papers, e.g., \cite{STEVE1996A}, as well as in computer animations available on internet \cite{STEVE1996B}.  The interesting feature, which was first observed with the  U-3/4Ti uranium alloy penetration data, is that while all available tungsten alloys retained a mushroomed head during the full penetration process, the U-3/4Ti uranium alloy penetrators formed a chiseled nose which resulted in the boring of a narrower channel which led to deeper penetration than tungsten \cite{MAGNE1990-}. Unfortunately this factual observation led to many simplified and exaggerated statements, which can even be found in the professional literature, that somehow tungsten penetrators have necessarily a ``large mushroomed head,'' while only uranium penetrators would have a ``self-sharpening nose.''

In reality, the 10 to 20\% penetration advantage discovered around 1975 with uranium alloys is due to a small effect that was apparently not properly understood until 1990, when Lee S. Magness and Thimoty G. Farrand published the correct explanation \cite{MAGNE1990-}.  According to them, the reason why U-3/4Ti uranium alloy penetrators keep a narrow profile during penetration is a metallurgical effect called ``adiabatic shear banding'' which implies that failures develop in the highly stressed nose region in such a way that the edges of the mushroomed head are quickly discarded, producing a sharpened or chiseled nose \cite{MAGNE1990-, MAGNE2001-}.  This shearing effect is small, as can be seen in shallow penetration simulations (e.g, \cite{STEVE1996A, STEVE1996B}), yet it is sufficient to produce the 10 to 20\% effect in deep penetrations which, according to numerous accounts, is claimed to be the main reason for deciding to produce large numbers of uranium penetrators since the late 1970s.

Soon after the origin of this technical advantage of uranium alloys was understood, and therefore shown not be due to any fundamental difference between uranium and other heavy materials, a considerable impetus was given to many research programs that were trying to find better alloys of tungsten and other materials (i.e., tantalum and hafnium) to replace and possibly outperform uranium alloys.  The realization that ``adiabatic shear behavior (...) aids the performance by minimizing the size of a mushroomed head on the eroding penetrator'' \cite[p.473]{MAGNE1990-}, along with ``concerns over environmental problems'' \cite[p.249]{ANDRE1992-} and depleted-uranium's ``drawbacks: health safety, cost of clean-up, environmental impact, political'' \cite[p.4,12]{DOWDI2000-}, all led to world-wide efforts to find substitute materials for uranium-based anti-tank weapons-alternatives that would be ``without the perceived hazards and political difficulties associated with DU'' \cite[p.1188]{MAGNE2001-}. To this effort anti-nuclear activists have substantially contributed by raising the public and parliamentary awareness of the problems associated with depleted-uranium.

For example, soon after the Gulf War, an ambitious program was initiated in South Korea, where the proximity of China (the world's largest producer of tungsten), the proximity of large potential markets such as Japan, and the uncertainties of its relations with North Korea, led to more than 13 patents and 23 scientific papers in less than ten years \cite{KADD-2001-, KIM--1998-, HONG2002-}.  In this, and other research programs, e.g., \cite{DOWDI2000-, MAGNE2001-, STEVE1998-}, the emphasis is on finding a tungsten or tantalum alloy, or an amorphous material based for example on hafnium, or even a more complex nanocrystalline material produced by nanotechnology \cite{MAGNE2001-}, that would exhibit adiabatic shear behavior and have the other required properties to be used for anti-tank penetrators.  Apparently, these programs are getting close to achieving their objectives \cite{DOWDI2000-, MAGNE2001-, HONG2002-}, and it remains to be seen whether these technical advances will translate into environmentally more acceptable anti-tank weapons.

However, while these technical developments are clearly important to improve one aspect of the terminal ballistic problem, one should not forget that they will potentially yield only a small contribution to the overall performance of an anti-tank system.  Moreover, there is an even more important consideration: does the perceived advantage of adiabatic shearing, namely a deeper but narrower penetration tunnel, really translate into a higher tank lethality?  The answer is almost certainly no!  The reason is that the motivation for seeking deeper but narrower penetration tunnels is the use of a single figure of merit, \emph{the penetration depth into semi-infinite RHA steel (penetration criterion)}, which does not take into consideration that the purpose of piercing armor is to inflict damage behind the armor: the penetration depth merely says that armor of such a thickness will just be perforated, with little energy left to inflict damage or loss of function behind the armor.

Therefore, a more realistic figure of merit is to take, for example, \emph{the complete perforation of a finite thickness of RHA steel plate (perforation criterion)}.  In effect, in that case some energy is left after perforation to inflict some damage behind the armor \cite[p.1160]{FARRA2001-}.  Using this simple change in the figure of merit, it was found that whereas using the traditional \emph{penetration criterion} uranium penetrators outperform tungsten penetrators by about 15 to 25\%, see \cite[Fig.2]{FARRA2001-}, using the \emph{perforation criterion} there is no longer any  significative difference between uranium and tungsten penetrators, see \cite[Fig.3]{FARRA2001-}.  The explanation of this important result is easily found, and related to the fact that tungsten penetrator tunnels are wider than uranium penetrator tunnels: ``uranium penetrates RHA much more effectively than tungsten; however, tungsten will produce a larger breakout effect.  Therefore, the two processes may cancel each other out, depending on the predominant failure mechanism of the target evaluated'' \cite[p.1163]{FARRA2001-}.  This result, obtained after a detailed discussion of the complicated problem of objectively defining the protection level (or seen from the other perspective the kill probability) of an armored vehicle, shows that serious methodological issues are still to be resolved by the designers of armored vehicles and anti-armor weapons!

Indeed, if the United States and a few other countries have proclaimed in the past 25 years that uranium-based anti-tanks were far better than other ones, several major countries, most prominently Germany,\footnote{But also Israel, China, Switzerland, Italy, Sweden, Spain, etc.} have not equipped their tank fleets with these, but instead have continuously improved the mechanical properties and the metallurgy of the tungsten alloy used in their anti-tank penetrators \cite{LANZ-2001-}.  And it turns out from what is known that these weapons are good enough to defeat the armor of any existing tank.  How is that possible?  A realistic criterion for objectively assessing the capabilities of these weapons is obviously the first step.  But there are other, more direct technical reasons.  For instance, some crucial mechanical properties which are important during the acceleration of the penetrator in a gun are significantly worse for uranium than for tungsten penetrators. ``DU penetrators therefore need stiffer and heavier sabots than tungsten rods, which compensates the slightly better impact behavior of DU in RHA'' \cite[p.1194]{LANZ-2001-}.  In a simpler language: what is gained by uranium alloys in the armor penetration phase, is lost in the initial launch phase and during the flight to the target.  But this does not close the discussion:  many more effects should be considered, especially in the interaction with complex multilayer tank shields \cite[p.255]{ANDRE1992-}, and even more so in the assessment of the lethality behind the armor \cite{FARRA2001-}.  However, as we have repeatedly stated, each of these effects has a small impact on the overall performance, so that whatever difference could exist between the properties of one or another heavy-alloy of tungsten or uranium, they are much more likely to cancel each other out than to cumulate and give a small advantage to either of them.

This is also true for one property that we have not yet considered, \emph{pyrophorism}:  Finely pulverized uranium spontaneously ignites in air.  This dramatic effect certainly adds a lethality component to uranium which tungsten, for example, does not have.  However, when a tank penetrator emerges behind an armor, it generates an intense shower of melted metallic fragments and particles which can badly burn the personnel, ignite the fuel-tank, or detonate the munition in the ammunition store. This is true whatever material the penetrator is made of, and it it likely that these break-out effects are more important for tungsten than for uranium penetrators.  Moreover, a modern main battle-tank may be equipped with a controlled atmosphere to protect against chemical weapons and avoid the propagation of fires: there could be little oxygen for uranium to burn!  Nevertheless, the pyrophoric nature of uranium adds a degree of lethality, which may increase the level of overkill of a weakly shielded armored vehicle defeated at short range, but not such a degree that it could significantly increase the kill radius of anti-tank rounds intended to defeat the best shielded main battle-tanks.

\subsection{Munitions for ground-attack aircrafts or close-in-defense guns}

In this last set of applications, the previous considerations about tank shields and kinetic-energy anti-tank weapons mean that the discussion can be quite short.

Ground-attack aircraft, such as the A-10 of the United States Air Force, have been designed and deployed ``to counter massive Soviet/Warsaw Pact armored formations spearheading an attack into NATO's Central Region (by firing armor piercing incendiary rounds) designed to blast through top armor of even the heaviest enemy tanks'' \cite{ROSTK2000-}.  There are thus two elements to assess: the armor-piercing and the incendiary capabilities ot the munition.  First, as is now amply demonstrated, the penetration superiority of uranium penetrators corresponds to only a (small) effect in the case of long-rod penetrators plunging into thick armor: this is not the case for ground-attack projectiles which are much less slender than long-rod penetrators, and which are intended to defeat the relatively thin top-armor of battle-tanks and the walls of relatively lightly armored vehicles such as personnel carriers.  Tungsten projectiles would be just as good.  As for the incendiary effect of A-10 projectiles due to the pyrophoric properties of uranium, it has not been possible to find any professional-level paper in which it is compared to the incendiary effect of tungsten shells containing an incendiary substance.  Nevertheless, since the pyrophoric property of uranium is not such that most of it will burn on impact, it can be inferred that a tungsten shell containing a sufficient amount of a truly effective incendiary substance could be just as devastating \cite{HURNI2003-}.

Finally,  it remains to discuss the use of heavy materials in munitions for close-in weapon systems, such as the U.S.\ Navy ``Phalanx'' designed to provide a last-ditch defense against sea-skimming missiles \cite{ROSTK2000-}.  In this application the main reason for using heavy metals is that a smaller and slimmer bullet has a significantly better external ballistics trajectory, which increases the precision of the system.  On the other hand, since sea-skimming missiles (or for that matter most types of precision-guided munitions and cruise missiles) are relatively ``soft'' targets, the armor penetration capabilities of heavy-metal alloys are of secondary importance.  The interesting aspect in the history of this application is that after deciding in 1978 to use a uranium alloy, the U.S.\ Navy decided in 1989 to change to tungsten alloys, ``based on live fire tests showing that tungsten met their performance requirements while offering reduced probabilities of radiation exposure and environmental impact''  \cite{ROSTK2000-}.  After this change from uranium to tungsten, further developments were made on the projectile, and a new tungsten alloy led to ``improved ballistic performance by 50\% compared to existing tungsten alloy penetrators'' \cite{CTC--2001-}.  While nothing is said about the reasons for this improvement, it most probably comes from the superior internal ballistics behavior of tungsten projectiles, which allows a tungsten bullet to be launched at a higher velocity than a uranium bullet.

\section{Are depleted-uranium weapons
         conventional or nuclear weapons?}

A central element in the depleted-uranium controversy is that on the one hand it is both radioactive and usable in civilian and military applications of nuclear energy, and that on the other hand the proponents of its ``non-nuclear'' uses claim that it can be treated as interchangeable with a heavy material such as tungsten. This leads to questions ranging from ``Is depleted-uranium a nuclear or conventional material?'' to ``Are depleted-uranium weapons conventional or nuclear weapons?' as well as to more fundamental questions such as ``What is a nuclear material?'' or ``What is a nuclear weapon?'' (And to their converse where the adjective nuclear is replaced by conventional, because ``conventional'' may not necessarily mean the same as ``non-nuclear''.)

These are perfectly reasonable questions, but they are very difficult to answer without taking many technical, legal, political, and even historical considerations into account.  A comprehensive analysis is not possible in a single paper, and certainly not in a paper written by a single author.  Nevertheless, it is possible to give a reasonable first answer to the question ``What is a nuclear material or weapon?'' and thus to develop useful pointers to answering the other questions.  Moreover, in relation to depleted-uranium weapons, it is possible to present some important elements, which have not been much discussed in the published literature, showing that their combat use cannot be separated from the potential use of nuclear weapons \cite{GSPON2002-}.  This is what will be done in the next three subsections.

\subsection{Nuclear and conventional materials}

Article XX of the Statute of the International Atomic Energy Agency (IAEA), which came into force in July 1957, defines a \emph{nuclear material} as being either a \emph{source material} or a \emph{special fissionable material}.  In simplified language, which hides many technical and legal subtleties, a \emph{special fissionable material} (e.g.,  uranium enriched in the isotopes 235 or 233, plutonium, etc.) is any fissionable material that has been artificially transformed or produced to make it more suitable for use in a nuclear reactor or a fission explosive; and a  \emph{source material} (e.g., natural uranium, uranium depleted in the isotope 235, thorium, etc.) is any material suitable for transformation into a special nuclear material (by enrichment, transmutation, etc.).  Therefore, the term \emph{nuclear material} covers all materials which may be used, either directly or after transformation, in a nuclear reactor or in the core of a \emph{fission} explosive, i.e., an ``atomic bomb,'' in which a self-sustaining chain reaction takes place.  Consequently, the term \emph{non-nuclear material} covers all other materials, even though they include light elements such as tritium and lithium which are the main source of the explosive power of a \emph{fusion} explosive, i.e., a ``hydrogen bomb.''   On the other hand, article XX makes no \emph{qualitative} distinction between natural-uranium and depleted-uranium, which are therefore both nuclear materials.

In practice, such definitions, which have the merit of being clear and consistent with both physical facts and common sense, were bound to create difficulties because of the dual-use characteristics, i.e.,  military and civilian, of nuclear materials.  Therefore, as soon as the IAEA tried to implement its role in the safeguard of nuclear materials, it had to introduce a quantitative element in order to establish a complicated accounting system for tracing the flow of nuclear materials throughout the world:  ``effective weights'' were assigned to nuclear materials on a subjective scale where, for example, 1 ton of depleted-uranium would be equivalent to only 50 grams of special fissionable material \cite{IAEA-1968-}, even though 1 ton of depleted-uranium could contain several kilograms of uranium 235.  On the other end of the spectrum, to deal with problems related to plutonium and highly enriched uranium, further definitions had to be introduced, e.g., the term \emph{nuclear-weapon-usable material} to designate nuclear materials that can be used for the manufacture of nuclear explosive components without transmutation or further enrichment.

The result was the beginning of a process which facilitated the non-military uses and the international trade of nuclear materials, while at the same time stretching out the scale used to classify these materials. Consequently, the materials sitting at either ends of the scale started to become exceptions of one type or another.  On the one end, nuclear-weapon-usable materials became more and more inaccessible to the non-nuclear-weapon states, with nuclear-weapon states exerting more and more direct pressure on those states, e.g., Yugoslavia, which had some amount of nuclear-weapon-usable materials on their territory \cite{STONE2002-}.  On the other end, the same nuclear-weapon states made a lot of efforts to trivialize the use of the main by-product of the nuclear industry: depleted-uranium.  This can clearly be seen in United States legislation \cite{AEC--1971, ISDCA1980-}, but also in international agreements such as the London Guidelines, which make exceptions for ``source material which the Government is satisfied is to be used only in non-nuclear activities, such as the production of alloys or ceramics'' \cite[Annex A]{LONDO1977-}.  In this guideline, ``Government'' refers to the exporting country, which means that such a government is authorized to allow the export of depleted-uranium for the purpose of making, for example, tank penetrators.

This ``stretching out'' process, which reflects the growing tension between nuclear-weapon states and non-nuclear-weapon states, or would-be-nuclear-wea\-pon states, is also visible in the evolution of concepts such as ``peaceful nuclear activities'' and ``proliferation-prone nuclear activities.''  While nuclear-weapon states insist on the peaceful and benign character of their broad and extensive nuclear activities, they impose very narrow and restrictive definitions of these activities by others.  This was particularly visible in 1991 when the IAEA was requested by the Security Council to carry out immediate on-site inspection in Iraq: for the first time, all activities prone to nuclear weapons proliferation had to be explicitly and comprehensively defined in an annex to a Security Council resolution so that the inspection team had a clear mandate to work from \cite[Appendix]{GSPON1996-}.  This annex confirmed IAEA's definitions of nuclear materials, but introduced other definitions and restrictions, such as the prohibition of the production of ``isotopes of plutonium, hydrogen, lithium, boron and uranium,'' which placed in the same category materials that are of importance to both nuclear and thermonuclear weapons \cite[p.11]{GSPON1996-}.

To conclude this subsection, it can be asserted that the concept of nuclear material as defined by the Statute of the IAEA is still valid today.  Therefore,   depleted-uranium is definitely a nuclear material according to international law, even though there is a trend to trivialize its use in so-called ``non-nuclear applications.''  This trivialization is clearly in contradiction with the legal status of the material, as well as with jurisprudence, because it is impossible to prove that a nuclear material is used  ``only in non-nuclear activities'' \cite{LONDO1977-} or ``solely to take advantage of (its) high density or pyrophoric characteristics''~\cite{ISDCA1980-}.  In fact, such ``declarations of intent'' are used as the main argument in the laborious argumentations put forward by those who argue that depleted-uranium weapons are neither radiological or nuclear weapons see, e.g., \cite{BEACH2001-}.\footnote{Similar kinds of  ``declarations of intent'' are used instead of factual evidence to assert that depleted-uranium weapons are neither incendiary or chemical weapons, despite the pyrophoric and toxic properties of uranium.}

\subsection{Nuclear and conventional weapons}

The words of most international arms-control agreements, such as the \emph{Nuclear Non-Proliferation Treaty (NPT)} and the \emph{Comprehensive Test Ban Treaty (CTBT)}, do not include definitions of essential terms, such as ``nuclear weapon'' or ``nuclear explosion,'' even though these terms refer to what is controlled or prohibited by the treaties.  The reason is the difficulty of defining them in unambiguous technical language that would be acceptable to all parties.  The definition and interpretation of these terms are therefore left to each individual party, which may share its interpretations confidentially with other parties, and possibly make them public in a declaration.

Consequently, in order to learn about what is really understood by the terms ``nuclear weapon'' and ``nuclear explosion'' one must analyse the declarations made by many countries, the statements they made during the negotiations, the opinion of leading technical and legal experts, etc.  Such an analysis is outlined in chapter 2 of reference \cite{GSPON1997-} where it is shown that the current practical definition of a ``nuclear explosion'' is much narrower that one would intuitively think.  In fact, what is strictly forbidden by the CTBT are not explosions in which any kind of nuclear reaction would produce some militarily useful explosive yield, but only explosions in which a diverging fission chain reaction takes place.  This is very far from the idea of a treaty stipulating ``not to carry out any nuclear test explosion or any other nuclear explosion'' (Article I of the CTBT) and leaves open the possibility of designing and testing new types of nuclear weapons in which no diverging fission chain reaction takes place!

In fact, this research is happening throughout the world, not just in the laboratories of the nuclear weapons states, but also in the major industrial powers such as Japan and Germany.  Using gigantic laser facilities, such the U.S.\ National Ignition Facility (NIF), France's Laser Mégajoule, Japan's GEKKO at Osaka, and similar tools in Germany and other countries, enormous progress is being made towards the design of small fusion bombs in full compliance with the CTBT --- which does not fully restrict the explosive use of nuclear fusion, unlike nuclear fission.  This activity is now the main advanced weapons research priority of the nuclear weapons laboratories,  which claim that its purpose is just the ``stewardship'' of stockpiled nuclear weapons. It is also the main route towards the controlled release of fusion energy in countries such Japan and Germany, which claim that these  small fusion explosives are only for peaceful applications.

As for defining the term ``nuclear weapon,'' the main source of difficulties is that the least unambiguous characteristics of the intuitive concept of a nuclear weapon are of a qualitative and subjective rather than quantitative and objective nature.  Indeed, it is the enormity of the difference in the destructive power of nuclear weapons compared to conventional weapons that has historically led to the qualitative distinction between ``nuclear'' and ``conventional'' weapons.  Similarly, it was their indiscriminate nature, and their targeting of cities rather than military forces, which led to the concept of ``weapons of mass destruction.''  Consequently, it is now difficult to argue that weapons made of a nuclear material like depleted-uranium are nuclear weapons, or even that new types of weapons in which a nuclear explosion takes place that is not forbidden by the CTBT are nuclear weapons.

It appears, therefore, that the most commonly accepted definition of a ``nuclear weapon'' --- that is of an enormously destructive weapon --- has created a situation in which less potent weapons based on similar materials or physical principles, and producing similar effects, but on a smaller scale, have become acceptable to governments as if they were ``conventional weapons.''

For example, when the British Under-secretary of state for defence, states that
\begin{quote}
``Nuclear, biological and chemical weapons are indiscriminate wea\-pons of mass destruction specifically designed to incapacitate or kill large numbers of people. DU ammunition is not'' \cite{MOONI2001-}
\end{quote}
he is right.  However, he is ignoring an important element that lies at the root of the ``dialogue of the deaf'' that characterizes the depleted-uranium debate, namely that depleted-uranium \emph{is} a nuclear material, that it \emph{is} radioactive, and that this \emph{has} important consequencies  --- especially in relation to nuclear weapons.  It can therefore be argued that depleted-uranium weapons are \emph{not} conventional weapons.

Similarly, the fact that the explosive yields of the laboratory-scale fusion explosions that will be produced in a few years, for example at the U.S.\ National Ignition Facility, will correspond to only a few kilogram equivalent of TNT  does not mean that these explosions are not nuclear explosions, which in fact they \emph{are} according to both physics and common sense.

\subsection{Depleted-uranium and fourth generation nuclear weapons}

As we have just seen, the huge difference in the damage produced by a single ``city-busting'' nuclear weapon compared to the impact of conventional weapons has led to the contemporary situation in which a radioactive material such as depleted-uranium is routinely used on the battlefield, and in which new types of nuclear explosives are being designed.  These nuclear explosives could have yields in the range of 1 to 100 ton equivalents of TNT, i.e., in the gap which today separates  conventional weapons from nuclear weapons, which have yields equivalent to thousands or millions of tons of TNT.

These new types of nuclear explosives are called fourth generation nuclear weapons (for an extensive discussion and a comprehensive bibliography see reference \cite{GSPON1997-}).  This name refers to a terminology in which the first generation corresponds to ``atomic'' or ``nuclear'' bombs, and the second to ``hydrogen'' or ``thermonuclear'' bombs.  The third generation corresponds then to the ``tailored'' or ``enhanced'' effects warheads such as the Enhanced Radiation Warhead (ERW, also called neutron bomb) which were never deployed in large numbers because they never found any truly convincing military use.  Moreover, since these third generation weapons still had the high yields and large radioactive fallouts characteristic of the first two generations they could not be used on the  battlefield as if they were some kind of trivial tactical weapons.

In comparison to the previous generations, fourth generation nuclear explosives are characterized by the following features :

\begin{itemize}

\item They will not contain a significant amount of nuclear-weapon-usable material (i.e., plutonium or highly enriched uranium) in which a self-sustaining diverging chain reaction could take place, so that their development and testing will not be forbidden by the  Comprehensive Test Ban Treaty;

\item They will produce relatively little radioactive fallout and residual radioactivity, because they will contain little or no fissile materials at all; 

\item They will derive the bulk of their explosive yield from fusion (which is why they may be qualified as ``pure fusion'' explosives) rather than from fission or other nuclear reactions, so that their radioactive effects will be those induced by fusion; 

\item They will have relatively low explosive yields, so that they will not qualify as weapons of \emph{mass} destruction.

\end{itemize}  

Consequently, compared to the third generation, fourth generation nuclear explosives are much more likely to find numerous military applications.  In particular, it is well known that the amount of conventional explosive that is delivered by precision-guided munitions like cruise missiles (i.e., 50 to 100 kg of chemical high explosives) is ridiculous in comparison to their cost:  some targets can only be destroyed by the expenditure of numerous delivery systems while a single one loaded with a more powerful warhead could be sufficient.  Therefore, the availability of fourth generation nuclear warheads with yields in the range of 1 to 100 ton equivalents of TNT will constitute a very dramatic change in warfare, a change that can only be compared to the first use of nuclear weapons in 1945 or the first deployment of intercontinental ballistic missiles in 1959.  Indeed, fourth generation nuclear weapons will have yields a thousand times larger than conventional weapons, that is a thousand time smaller than nuclear weapons of the previous generations, which means that they are much more likely to be used on the battlefield than first, second, or third generations nuclear weapons, especially since they correspond to a strongly perceived military need.

In practice, while proponents of fourth generation nuclear weapons may convincingly argue that just like depleted-uranium weapons they are not weapons of mass destruction, that they can be used in discriminate ways, etc., it will be more difficult for them to dismiss the fact that they nevertheless produce effects that do not exist with conventional explosives:  namely an intense burst of radiation during the explosion, and some residual radioactivity afterwards. 
 
The proponents will therefore argue that the burst of radiation (mostly high-energy fusion neutrons) will affect only the target area, and that the only long-term collateral damage will be the ``low'' residual radioactivity due to the dispersal of the unburnt fusion fuel and to a smaller extent to the interaction of the neutrons with the ground, the air, and the materials close to the point of explosion...

It is at this stage of the critical assessment of the military consequences of the potential use of fourth generation nuclear weapons that an important finding was made: \emph{The expenditure of many tons of depleted-uranium has a radiological impact comparable to that of the combat use of many kilotons TNT equivalent of pure-fusion nuclear explosives} \cite{GSPON2002-}.

This finding, which is reported in the Proceedings of the Fourth International Conference of the Yugoslav Nuclear Society, Belgrade, Sep.30 - Oct.4, 2002, was very surprising at first.  This is because it was totally unexpected that the radiological impact of the battlefield use of depleted-uranium could be compared to that of the battlefield use of a large number of hypothetical fourth generation nuclear warheads.  But this is what the laws of physics and what is known about the effects of nuclear radiations on human beings imply:  the use of depleted-uranium weapons creates a residual radioactive environment that provides a yardstick which can be used by the proponents of fourth generation nuclear weapons to demonstrate that the radioactive burden due to their use is ``acceptable.''

For instance, it is found in reference  \cite{GSPON2002-} that the expenditure of one ton of depleted-uranium in the form of bullets has a long-term radiological impact equivalent to the use of many kilotons of hypothetical pure-fusion weapons.  This means that between 100 and 1000 precision-guided munitions (which today deliver only 10 to 100 kilograms of high-explosives), each carrying a pure-fusion warhead with a yield of 1 to 10 tons of high-explosive equivalent, could be used to produce a similar radiological impact.  Since about 400 tons of depleted-uranium were used in Iraq, and about 40 in Yugoslavia, the radiological impact in these countries corresponds to that of using tenths of thousands of precision-guided munitions tipped with fourth generation nuclear warheads, i.e., many more precision-guided delivery systems than were actually used in these countries.

Thus, while depleted-uranium weapons create a low radioactive environment, this radioactive burden is not less negligible than the predictable radiological impact of the new types of nuclear weapons that are under development.  The battlefield use of depleted-uranium has therefore created a military and legal precedent for the use of nuclear weapons which produce a radioactive burden that is much less than that of existing types of nuclear weapons, but which have a destructive power about a thousand times larger than conventional explosives.

\section{Discussion and conclusion}

In this paper two important new contributions have been made to the depleted-uranium debate:  (1) a critical appraisal of the conventional weapon's use of heavy materials showing that depleted-uranium alloys have no truly significant technical or military advantage over tungsten alloys for that purpose; and (2) an analysis showing that depleted-uranium weapons cannot be classed as conventional weapons and that they belong to the diffuse category of low-radiological-impact nuclear weapons to which emerging types of nuclear explosives also belong.

These contributions serve to complete and reinforce the numerous arguments that have been put forward to show that depleted-uranium weapons are illegal according to international law and contrary to the rules of war.  They also clarify the discussion of the reasons why depleted-uranium weapons have been made and why they were used in Iraq and Yugoslavia. 

Since the main battlefield use of depleted-uranium was in anti-tank weapons, let us review and critique six of the more obvious military reasons that must have played a role in deciding to use them against Iraqi and Serbian armor:

\begin{enumerate}

\item \emph{The most effective ammunition available had to be used in order to get quick results.}  Depleted-uranium munitions are at best only marginally more effective than tungsten munitions.  Moreover,  the majority of Iraqi tanks were not of the latest Soviet generation (such as the T-80), and the use of depleted-uranium rounds in Yugoslavia against lightly armored tanks and personnel carriers was a gross overkill.  In fact, just as effective (and less costly) non-radioactive anti-tank weapons were available in large numbers during both the Iraqi and Yugoslavia campaigns.

\item \emph{Depleted-uranium munitions have the additional advantage that uranium is pyrophoric.}  The incendiary effect of depleted-uranium anti-tank rounds is only a marginal contribution to their tank lethality effect.  On the other hand, the pyrophoric property of uranium is the \emph{main} short-term cause of the dispersion of uranium in the environment. 

\item  \emph{Depleted-uranium munitions were available but had never been tested on the battlefield.}  Their use in 1991 broke a 46-year-long taboo against the intentional battlefield use of radioactive materials.  It is therefore particularly shocking that during the Gulf War the British Ministry of Defence quickly adapted depleted-uranium ammunition developed for the new Challenger-2 tank so as to use it on the existing Challenger-1 tank fleet \cite{BEACH2001-}. This meant that the munition could be labeled ``combat proven'' after the war.

\item  \emph{Depleted-uranium is a radioactive material which inspires pride in its users, fear in its victims, and strong reactions in the bystanders.}  Indeed, depleted-uranium long-rod penetrators are the ``favorite rounds'' of tank gunners and the ``least acceptable rounds'' to a large share of the public opinion and environmental activists.  Consequently, the numerous and often excessive reactions of the anti-depleted-uranium activists unwillingly contributed to the military propaganda machine by amplifying the military-psychological advantage of using depleted-uranium instead of a non-radioactive material.

\item  \emph{Depleted-uranium weapons have an overall long-term radiological effect comparable to that of pure-fusion nuclear weapons.}  Like the previous reason, there is no direct evidence that this reason was taken into account at some stage of the decision making process.  However, the analysis made in reference \cite{GSPON2002-}, as well as considerable indirect evidence,\footnote{For example, the care that is taken in all published documents to avoid any open discussion of the long-term radiological impact of pure-fusion nuclear weapons: see, e.g., reference 6 cited in \cite{GSPON2002-}. In fact this impact is very simple to calculate, and immediately suggests a comparison with low-radioactive contamination such as arising from the combat use of depleted-uranium \cite{GSPON2002-}.} show that the military planners\footnote{Here, ``military planners'' refers not so much to the high-ranking soliders who planned the operations in Iraq or Yugoslavia (or those who previously approved the introduction of depleted-uranium weapons into arsenals) but rather to the analysts in defense departments and laboratories, as well as in military think-tanks and universities, who shape general policies according to both technical and political considerations, either by direct input to the decision makers, or by omission.} are acutely aware of the \emph{full} consequences of using depleted-uranium and the lessons that can be learned from its use.  In particular, it is indisputable that the use of a radioactive material in Iraq and Yugoslavia has created a military and legal precedent.  Similarly, it is  indisputable that this use has provided a first test of the acceptability of future weapons that would produce a low level of radioactivity.  In fact, the use of depleted-uranium weapons has proved to be acceptable, both from a military point of view because the induced radioactivity did not impair further military action, and from a political standpoint because most political leaders, and shapers of public opinion, did not object to their battlefield use. 

\item  \emph{Depleted-uranium weapons are less expensive than tungsten weapons.}  This reason has been left for the end because it is not strictly-speaking a military reason and because it is fallacious in two respects.  First, it is clear that the cost of a single anti-tank round that is capable of destroying, or saving, a main battle tank is marginal in comparison to the cost of such a tank.  Therefore, using a phrase that is much used in such a context, if ``gold plating'' would improve the performance of anti-tank rounds, they would all be gold plated!  Consequently, it can be asserted that even if depleted-uranium was more expensive than tungsten and depleted-uranium rounds were only marginally better than tungsten rounds, uranium rounds would still be preferred by the military.  Second, it turns out that the market value of depleted uranium is artificially low because it is a surplus of the nuclear industry.  However, this low price of the raw material proves to be misleading: the additional costs incurred by choosing uranium (processing difficulties, problems due to pyrophorism and radioactivity, clean-up of the environment after use, etc.) mean that the full \emph{life-cycle cost} of uranium is much higher than that of tungsten \cite{ROUSS2001-}.  Therefore, if the full life-cycle cost is taken into account, the preference easily goes to tungsten, an observation that the tungsten industry has not failed to make \cite{ROSKI2001-}.  To sum up, in a situation such as the 1991 Gulf War where a full range of antitank weapons was available --- including a number of tungsten antitank weapons, which it would have been opportune to compare in combat to uranium ones --- the price argument was truly of secondary importance.

\end{enumerate}

Reflecting on these military reasons, one sees that they basically take the same line of argument that is always developed in order to justify the introduction of some new controversial weapon.  In the case at hand, there are even strong echos of the ``neutron bomb debate'' of the mid 1980s \cite{GSPON1984-, SAHIN1985-, HARRI1986-}.  In that debate, the proponents presented the neutron bomb as a kind of almost magical superweapon that would defeat a Soviet tank invasion without producing large collateral damage to the surrounding towns and cities.  The argument was shown to be plain wrong by simply making an analysis of modern tank armor technology and by calculating the protection that would be provided by such armor against the neutrons from a neutron bomb \cite{SAHIN1985-}.  Similar analyses were made in section 2 of this paper (where a rather long argument was required to complete the proof that depleted-uranium rounds are no ``silver bullets'') and in reference \cite{GSPON2002-}, using only published professional papers, and purely scientific arguments.  The conclusions are that depleted-uranium weapons are not superweapons of any kind, and that their perceived military advantages, as well as their perceived environmental disadvantages, are vastly exaggerated by both the proponents and the opponents of these weapons.

In the case of the neutron bomb debate, it turned out that rational and measured arguments of the kind put forward in references \cite{GSPON1984-, SAHIN1985-, HARRI1986-} had a remarkable impact on the outcome of the debate, which turned to the advantage of the opponents.  It is therefore of interest to see what such arguments can say about the possible reasons \emph{why} depleted uranium weapons have been put into arsenals despite the numerous environmental, medical, technical, legal, and political objections that have been anticipated, and are still raised against their fabrication, deployment, and use.

First of all, as can be seen in the professional literature, of which the bibliography of this paper only gives a limited overview, there has been a lively debate between ballistic weaponeers about the merits and demerits of depleted-uranium versus tungsten projectiles.  This debate has been running for fifty years now, and the use of depleted-uranium in Iraq and Yugoslavia has only reinforced the convictions of the weaponeers opposed to  depleted-uranium.  In fact, while the U.S.\ Navy had already reverted to tungsten years before the Gulf War, and the U.S.\ Air Force has now decided that its next generation tank killer will not use depleted-uranium \cite{FIALKA2003B}, no country that had decided against depleted-uranium penetrators changed its policy after 1991.\footnote{The main reason given by these countries has always been that it is contradictory to impose almost zero-level release of radioactivity to the environment by the nuclear industry, while at the same time to accept its uncontrolled release by the military.}  

Secondly, there are the well-known facts that the price of depleted uranium has been made artificially low in order to encourage its use \cite{DOLE-1978-}, and that the price of tungsten (like that of other strategic materials) is mainly a function of political decisions such as the level of releases from the huge stockpiles maintained in Russia and the United States \cite{ROSKI2001-}.  Moreover, there is the observation that the unclassified American technical literature is clearly biased in that it mostly refers to the performance of anti-tank \emph{penetrators} rather than to the performance of complete anti-tank weapon \emph{systems} (see also \cite{ROSTK2000-}).  All of this suggests that the push towards the use of depleted-uranium has been much stronger than was ever justified by a small perceived advantage of uranium alloys over tungsten alloys in one specific application: long-rod penetration into homogeneous armor.

Thirdly, there is a long history of events showing that the ``nuclear lobby'' (which in the 1950s and 60s was the incarnation of ``progress and modernity'') has always had a very strong influence on the course of economic and military affairs in all the nuclear powers.  In particular, one can see its influence in the politics behind the development of new nuclear weapons, for example the neutron bomb, or new sources of nuclear energy, for example micro-explosion fusion.  It is tempting to speculate that a policy which consists of making the price of a material like depleted-uranium as low as possible is both an investment into the future and a way to trivialize the use of radioactive materials in all circumstances. 

Finally, there is the very grave military and legal precedent created by the combat use of a radioactive material, which is a clear violation of the spirit, if not the letter, of a norm that was in force since 1945.  To argue that nobody holding a position of competence or responsibility was aware of the full consequences of breaking this norm would be quite unreasonable.  On the contrary, considering the existence of a lively internal debate on the consequencies of the battlefield use of depleted-uranium and other nuclear materials, it is certain that these had been thoroughly investigated long before the 1991.  One must admit, therefore, that the choice made in favor of using depleted-uranium took into account the fact that its battlefield use would trivialize the military use of radioactive materials, and would therefore make the use of nuclear weapons more probable.

In conclusion, it can be argued that besides its military function, the use of depleted-uranium in Iraq and Yugoslavia must have served a political purpose: to prepare for the progressive introduction of fourth generation nuclear weapons whose battlefield use will cause a low (but non-negligible) residual radioactive environment.  It may even be possible to argue that depleted-uranium was used in Iraq --- and then in Yugoslavia where there was little military reasons for using it --- in order to test the opposition of the Western public opinion to the induction of radioactivity on the battlefield, and to get the world population accustomed to the combat use of depleted-uranium and fourth-generation nuclear weapons.

\section{Acknowledgments}

Work on this report has benefited from discussions and correspondence with many people. I wish to thank in particular the following persons for their valuable contributions: Bernard Anet, Hugh Beach, Thomas Cochrane, Jean-Pierre Hurni, Willi Lanz, Peter Low, and Bruno Vitale.

\end{document}